\documentclass{article}
\usepackage{graphicx}%
\usepackage{multirow}%
\usepackage{amsmath,amssymb,amsfonts}%
\usepackage{amsthm}%
\usepackage{mathrsfs}%
\usepackage[title]{appendix}%
\usepackage{xcolor}%
\usepackage{textcomp}%
\usepackage{manyfoot}%
\usepackage{booktabs}%
\usepackage{algorithm}%
\usepackage{algorithmicx}%
\usepackage{algpseudocode}%
\usepackage{listings}%
\usepackage{bbold}
\usepackage{amsmath}
\usepackage{amsthm}
\usepackage{amssymb}
\usepackage{amsfonts}
\usepackage{tensor}
\usepackage{braket}
\usepackage{slashed}

\theoremstyle{definition}

\newcommand{\vect}{\textbf}

\newcommand{\x}{\vect{x}}

\newcommand{\vd}{\boldsymbol{\partial}}

\begin{document}
	\title{Phase Space structure on Clifford Algebras}
	\author{Calum Robson\footnote{c.j.robson@lse.ac.uk}}
	\maketitle
\begin{abstract}
	I argue that the Hodge structure on a Euclidean Clifford algebra $Cl(n)$ provides a way to generalise Kähler structure to higher dimensions, in the sense that the paired variables are now associated with $k-$ and $(n-k)-$ dimensional subspaces rather than with vectors. This puts a phase space structure on Clifford algebras, and so allows us to construct a Hamiltonian dynamics on these multilinear variables. This construction shows that alternating pairs of subspaces obey commuting and anticommuting dynamics, hinting that this construction is indeed a natural one, with interesting new behaviour. 
\end{abstract}
	\section{Introduction}
	The aim of this paper is to investigate how the Kähler structure usually associated with complex manifolds can be generalised to Clifford algebras. This is inspired by the programme of Clifford Analysis \cite{Delanghe1992} \cite{Roberts2022} which extends many results of complex analysis-- for example the Cauchy integral formula-- to higher dimensions via Clifford algebras. This paper is a first step, and a very preliminary one, towards applying this approach to the theory of complex manifolds \cite{Huybrechts2005}. \\
	I will begin by giving an overview of Kähler structure and its relation to Physics-- in particular it is important for defining Hamiltonian dynamics on phase spaces. I will then discuss Clifford algebras, and the Geometric Algebra  representation used in this paper. Finally I will motivate the search for for a phase space structure on Clifford algebras by discussing the, `Stabilised Poincare-Heisenberg Algebra' (SPHA), which is isomorphic (depending on the choice of metric) to either the Clifford algebra $Cl(3,1)$ or $Cl(4)$, and includes the Heisenberg algebra as a subalgebra. 	None of this material is original. For much more detail about Kähler structures, see, e.g.  \cite{Huybrechts2005}\cite{Ballman2006} For more information on Clifford algebras see the references \cite{Lounesto2009}\cite{Doran2003}\cite{Macdonald2010}\cite{Lengyel2024}. The papers on the SPHA are cited in the text. \\
	The original argument of this paper is that the Hodge relation \cite{Schwarz1991}
	\begin{equation}
		g(A,B)\mathcal{I}=A\wedge\star B
	\end{equation}
defined on a Clifford algebra gives a higher dimensional version of Kähler structure. After discussing the logic behind this suggestion, I will provide evidence that it is a sensible one  by using it to give a phase space structure to Clifford Algebras, and a corresponding Hamiltonian dynamics. The generalisation is that for Clifford algebras, the dual positions and momenta are orthogonal $k-$ and $(n-k)$-- dimensional subspaces of an $n$ dimensional vector space, rather than orthogonal vectors. \\
Kähler geometry is only defined for manifolds with a positive definite metric (see \cite{Flaherty1976} for a discussion of the issues involved). Therefore I will only consider Euclidean Clifford algebras. I hope to discuss the non-Euclidean case in a future paper. 
	\subsection{What is Kähler Structure?}
 A Kähler Structure is a triple, $\big\{\omega, g, J\big\}$  on a $2n$ dimensional real vector space (or an $n$ dimensional complex vector space) , where $\omega$ is an antisymmetric two-form, $g$ is a metric, and $J$ is a complex structure satisfying $J^{2}=-1$; and where this triple is related by
\begin{equation}\label{eqn:Kähler}
	g(u,v)=\omega(u,Jv)
\end{equation} 
We have two mappings from a vector space $\mathcal{V}$ to its dual space $\mathcal{V}^{\star}$. These two mappings are defined by the metric $g(u,v)$ and the antisymmetric form $\omega(u,v)$. These mappings are 
\begin{align}
	u^{\star}_{1}&=g(u, \cdot)\\ \nonumber
	u^{\star}_{2}&=\omega(u, \cdot)
\end{align}
A space is Kähler if these two mappings are related by
\begin{equation}
	g(u,v)=\omega(u, Jv)
\end{equation}
Where $J$ is a complex structure, satisfying $J^{2}=-1$. This means that the mappings are orthogonal, in the sense that the vectors $v$ and $Jv$ on which they are equal are orthogonal, since: 
\begin{equation}\label{eqn:Kählerorth}
	g(v, Jv)=\omega(v, J(Jv))=\omega(v, J^{2}v)=\omega(v, -v)=-\omega(v, v)=0
\end{equation}
since $\omega$ is antisymmetric. On the complex plane, multiplication by $i$ rotates through $\pi/4$, so $z$ and $iz$ are orthogonal. The complex structure $J$ also rotates $v$ to an orthogonal direction $Jv$. We also require that 
\begin{equation}
	g(v, Jw)=g(J^\star v, w)
\end{equation}
so that
\begin{equation}
\lvert Jv\rvert^{2}=g(Jv, Jv)=g(J^{\star}Jv, v)=g(v,v)=\lvert v\rvert^{2}
\end{equation}
and therefore the magnitudes of $v$ and $Jv$ are the same. If our vector space $\mathcal{V}$ is complex then this follows from the condition that $g$ is a hermitian metric. We can always find such a structure locally, e.g. by taking $J$ to be a suitable rotation within the tangent space at a point-- this is called an \textsl{almost Kähler structure}. For a manifold to have Kähler structure then this must hold globally (the complex structure $J$ must be integrable), which is a very strong condition. \\
The antisymmetric $\omega$ is called a symplectic form, and is related to Hamiltonian mechanics.  Consider a function $H(p,q)$ which must be constant under the evolution of some parameter $t$ which I shall refer to as the time parameter . If $H$ has no explicit time dependence then we have
\begin{equation}
	\frac{dH}{dt}=\frac{\partial H}{\partial p}\frac{dp}{dt} +\frac{\partial H}{\partial q}\frac{dq}{dt}
\end{equation}
We can always set this to zero provided that
\begin{equation}
	\dot{q}\equiv\frac{dq}{dt}=\frac{\partial H}{\partial p}; \ \dot{p}\equiv\frac{dp}{dt}=-\frac{\partial H}{\partial x}
\end{equation}
Which are precisely Hamilton's equations. This argument generalises to more than one pair of conjugate variables, on the assumption that the variations in each $(q_{i},p_{i})$ pair can only cancel one another, and not the variations of other variables. It is possible to define a, `multisymplectic geometry' \cite{Ryvkin2019} where more than just pairwise variations cancel but this will not concern us in this paper. 
Suppose we now want to calculate the time variation of another function $G(t,q,p)$ which may be explicitly dependent on time
\begin{align}\nonumber
	\frac{dG}{dt}=\frac{\partial G}{\partial q}\frac{dq}{dt}+\frac{\partial G}{\partial p}\frac{dp}{dt}+\frac{\partial G}{\partial t}\\ \nonumber
	=\frac{\partial G}{\partial q}\frac{\partial H}{\partial p}-\frac{\partial G}{\partial p}\frac{\partial H}{\partial q}+\frac{\partial G}{\partial t}\\
	\equiv\big\{G,H\big\}+\frac{\partial G}{\partial t}
\end{align}
where
\begin{equation}
\big\{f,g\big\}=\frac{\partial f}{\partial q}\frac{\partial g}{\partial p}-\frac{\partial f}{\partial p}\frac{\partial g}{\partial q}
\end{equation}
is the Poisson Bracket. Hence if we assume the quantity $H$ is conserved, we can define the time evolution of all other quantities in terms of the derivatives of $H$. \\
How is this related to the symplectic form $\omega$? We can show that for a suitable choice of basis, the Darboux basis $\{ x_{i}, p_{i}\}$, we have
\begin{equation}\label{eqn:poissonomega}
\{f,g\}^{-1}=\{f,g\}^{T}=\omega\big( \nabla f, \nabla g \big)
\end{equation}
This allows us to define the dynamics of the Hamiltonian function via 
\begin{equation}
	\omega(V_{H},  \cdot)=-D_{p}H(\cdot)
\end{equation}
where $V_{H}$ is called the Hamiltonian vector field. Assuming we have a Kähler structure then
\begin{equation}
	\omega(V_{H}, V_{H})=0=-D_{p}H(V_{H})=V_{H}\cdot\nabla H
=	g(V_{H}, J^{-1}V_{H})
\end{equation}
which shows that $V_{H}$ is everywhere orthogonal to $\nabla_{H}$ and so is the tangent vector to contours of $H$. Therefore $H$ is conserved as its derivative along its own tangent vectors is zero. 
	\subsection{Geometric Clifford Algebras}
	Since this paper is about extending the ideas behind the Kähler structure to Clifford algebras, I will now give a brief overview of these objects, focussing on the aspects important for the later results. Given an $n$-dimensional vector space $\mathcal{V}$, an orthonormal basis $\{e_{i}\}$ for this space, and a metric $g(e_{i}, e_{j})=g_{ij}$, we can associate a Clifford Algebra to this space by keeping the vector addition and scalar multiplication from $\mathbb{V}$, and adding a vector multiplication satisfying
	\begin{equation}\label{eqn:Clifforddef}
		e_{i}e_{j}+e_{j}e_{i}=2g_{ij}
	\end{equation}
We are here only focussing on Clifford algebra s $Cl(n)$ where this metric is Euclidean (i.e. positive definite). Note that equation (\ref{eqn:Clifforddef}) implies that the products of distinct $e_{i}$ form the standard exterior algebra, so if $e_{i}$ is a vector, $e_{1}e_{2}$ is a bivector and $e_{1}e_{2}e_{3}$ is a trivector, and so on. We call the number of basis elements which multiply to make a particular element the $\textsl{grade}$ of an element, and the highest grade element (of grade n, the exterior product of all the basis vectors) is called the pseudoscalar, often denoted $I$. The simplest example of this construction are the complex numbers $\mathbb{C}$. We can construct these in two ways; first of all as the Clifford algebra 
\begin{equation}
	Cl(1)=\{1, e_{1}\}
\end{equation}
generated from $\mathbb{R}$, with $e_{1}^{2}=-1$. Another (and more commonly seen) construction is to start from the vector space $\mathbb{R}^{2}$, with the standard basis $e_{1}, e_{2}$, and the standard euclidean metric $g(e_{i}, e_{j})=\delta_{ij}$. Then we get the algebra 
\begin{equation}
	Cl(2)=\big\{1, e_{1}, e_{2}, e_{1}e_{2}\big\}
\end{equation}
Here, $e_{1}e_{2}\equiv I$ is the pseudoscalar, and satisfies $ I^{2}=-1$.  The algebra generated by the even grade part of $CL(2)$, called $Spin(2)=\{1, I\}$ is isomorphic to $C$. In this paper, I will adopt the perspective that Clifford algebras should be taken as higher dimensional extensions of the Complex Numbers, insofar as the Clifford algebra $Cl(n)$ is the algebra canonically associated to $\mathbb{R}^{n}$. \\
There are many ways to visualise or represent the complex numbers (for example as the quotient $\mathbb{P}(x)/(x^{2}+1)$, where $\mathbb{P}(x)$ is the set of all polynomials. Often the most convenient way to approach the complex numbers is to represent them geometrically via the complex plane or Argand diagram. Similarly, whilst there are many ways to think about Clifford algebras, it can often be convenient to adopt a geometric representation. There are many alternative representations, but the most basic is to regard the vectors $e_{i}$ as unit lines, the bivectors $e_{i}e_{j}$ as unit areas, the trivectors as volumes-- and so on. This is the approach I shall take in this paper. Such a representation is known as a Geometric Algebra.  \\
We refer to a Clifford Algebra element consisting only of linear combinations of elements of a single grade as a $\textsl{blade}$.  I shall refer to a blade which is a product of $k$ basis vectors as a a basis blade, and denote it $e_{K}$. For example, $e_{1}e_{2}e_{3}$ is a basis blade, but $e_{1}e_{2}e_{3}+2e_{2}e_{3}e_{4}$ is a grade 3 blade, but not a basis blade. A linear sum of blades is called a multivector. \\
We often denote a blade of grade $k$ as $\langle A\rangle_{k}$. When we multiply two blades  $\langle A \rangle_{i}$ and $\langle B \rangle_{j}$, the lowest grade object we can form has grade $\lvert i-j\rvert$. It is called the dot product,  $A\cdot B$ or $\langle A\cdot B\rangle_{i-j}$. The highest grade element has grade $i+j$, and is called the wedge product, denoted $A\wedge B$ or $\langle A\wedge B\rangle_{i+j}$. The set of basis vectors $\{e_{i}\}$ equipped with the wedge product forms the standard exterior algebra, and the Clifford algebra relation (\ref{eqn:Clifforddef}) can be written as $e_{i}\cdot e_{j}=g_{ij}$. \\
A key algebra for physics which will be referred to in the next section is the Spacetime Algebra, $Cl(3,1)$, sometimes called the Spacetime algebra. It is the geometric algebra generated by Minkowski space $\mathbb{R}^3,1$. The notation here means that three variables square to 1, and one variable ($e_{0}$) squares to $-1$. t is basis given by
\begin{equation}
	Cl(3,1)=\Big\{ 1, \gamma_{a}, \gamma_{0}\gamma_{i}, \gamma_{i}\gamma_{j}, \gamma_{a}\gamma_{b}\gamma_{c}, I \Big\}
\end{equation}
Here $\{a,b,c \}$ run from 0 to 4, whereas $\{i,j,k\}$ refer to the spacelike directions, $\{1,2,3\}$. This basis for the spacetime algebra contains four vectors $\gamma_{a}$, three timelike rotations (boosts) $\gamma_{0}\gamma_{i}$; three spacelike rotations $\gamma_{i}\gamma_{j}$; four $3-$volumes $\gamma_{a}\gamma_{b}\gamma_{c}$; and the volume form $I\equiv\gamma_{0}\gamma_{1}\gamma_{2}\gamma_{3}$. This is a pseudoscalar of grade 4. The Euclidean Clifford algebra $Cl(4)$ has exactly the same basis but with all four basis vectors squaring to 1.  \\
An important concept for Clifford algebras is duality. There has been some confusion regarding the best way to define this, and I am here largely following Leyngel's approach \cite{Lengyel2024}. First note we can define an involution $\dagger$ on a Clifford algebra, which reverses the order of the basis elements making up an element. It is defined via
\begin{equation}
(e_{1}e_{2}...e_{k})^{\dagger}=(e_{k}...e_{2}e_{1})
\end{equation}
and is called the $\textsl{reversion}$. Geometrically, it swaps the chirality of the $A$. We can use this to define different sorts of duality. First of all, note (e.g.from the singular values decomposition) that the adjoint of a linear transformation gives the opposite rotation and same dilation, whereas the inverse gives the opposite rotation and the inverse dilation. We can define these on Clifford algebras, where the inverse and adjoint of a grade $k$ blade $A$ are both grade $k$. They both have the opposite chirality to $A$, but the adjoint has the same magnitude and the inverse has inverse magnitude. We can also define two dual operators, the geometric inverse and the Hodge dual of $A$ , which are grade $n-k$ and occupy the $n-k$ directions orthogonal to $A$, as follows:
	\begin{center}
		\begin{tabular}{c | c}
			\textbf{Scalar} 	\textsl{(k dim.)} &\textbf{ Pseudoscalar (Volume form)} (\textsl{(n-k) dim.)}\\
			Inverse 	$A^{-1}=\frac{A^{\dagger}}{\lvert A\rvert^{2}}$ & Geometric inverse $A^{-1}\mathcal{I}$\\
			Adjoint $A^{\dagger}$ & Hodge Dual  $\star A =A^{\dagger}\mathcal{I}$
		\end{tabular}
	\end{center}
The Hodge dual corresponds to the adjoint in the sense that it has the same magnitude, and the inverse corresponds to the geometric inverse. We can also use $\dagger$ to define the inner product $g(A,B)=B^\dagger\cdot  A$. With this definition we can write the Hodge relation between two blades as  
\begin{equation}
g(A,B)\mathcal{I}=(B^{\dagger}\cdot A)\mathcal{I}=A\wedge\star B=A\wedge(B^\dagger I)
\end{equation}
This will be very important later. The norm of a unit basis blade is therefore given by $e_{K}^{\dagger}e_{K}$. This is always 1 in the Euclidean case. Note that this implies that for the pseudoscalar,  $I^{\dagger}I=1$ for Euclidean metrics. We can extend all these definitions to general Clifford algebra elements (multivectors) by taking linear combinations of blades. The final thing to introduce is the vector derivative, or Dirac operator
\begin{equation}\label{eqn:vecder}
	\vd=\frac{\partial}{\partial\x}\equiv \ e_{i}\frac{\partial}{\partial x}\\
\end{equation}
Applying the vector derivative to a blade will either raise or lower the grades of the components of multivector by 1. In general it will do both. We therefore write
\begin{equation}
	\vd\vect{z}=\vd\cdot\vect{z}+\vd\wedge\vect{z}
\end{equation}
where  $\vd\cdot\vect{z}$ and $\vd\wedge\vect{z}$ are the grade lowering and grade raising parts respectively. \\
For a non-orthonormal basis $\{f_{i}\}$ we would have $\vd=f_{i}^{-1}\partial_{x_{i}}$. Later in this paper I will introduce a multivector derivative-- for a unit k-  blade $e_{K}$ associated to a variable $z_{k}$, where $Z_{k}=z_{k}e_{K}$. I will define it as 
\begin{equation}
\vd_{Z_{k}}=e_{K}^{\dagger}\partial_{z_{k}}
\end{equation}
Since $e_{K}$ is a unit blade, $e_{K}^{\dagger}=e_{K}^{-1}$. I have used this definition for consistency with the non- orthonormal case. When $k=1$ this has no effect (hence why it does not appear for the standard vector derivative), but it will make a difference for higher grades. 
\subsection{The Stabilised Poincare-Heisenberg Algebra}
One motivation for the construction in this paper comes from an area of pure mathematics called, `Lie algebra stability theory'. What does it mean for a Lie algebra to be stable? The technical machinery is complicated, but the idea is that if we start with the Lie algebra defined by
	\begin{equation}\label{eqn:liealg1}
[f_{a}, f_{b}]=C_{ab}^{c}f_{c}
	\end{equation}
where the $f_{a}$ are the lie algebra generators, and the $C_{ab}^{c}$ are structure constants, and deform the algebra to the one defined by
	\begin{equation}\label{eqn:liealg2}
	[f_{a}, f_{b}]=\big(C_{ab}^{c}+\lambda_{ab}^{c}\big)f_{c}
\end{equation}
where we have introduced a variation $\lambda_{ab}^{c}$ into the structure constants. We can now ask whether the deformed algebra defined by (\ref{eqn:liealg2}) is isomorphic as a lie algebra to that defined by (\ref{eqn:liealg1}). If they are isomorphic, for all sufficiently small $\lambda_{ab}^{c}$, we say the original lie algebra (\ref{eqn:liealg1}) is stable. If not we say it is unstable. \\
One way to think of this is to imagine we have a moduli space of Lie Algebras, with points labelled by the structure constants $C_{ab}^{c}$. We can define a flow on this moduli space by varying the structure constants, and then we call a Lie Algebra, `stable' if it is a stable fixed point on this flow. The full details of this construction are considerably more complicated, however, and rely on the cohomology of this moduli space. More information can be found in \cite{Nijenhuis1967}. \\ 
The relevance of this to physics is the fact that physical theories seem to use stable algebras. This makes sense-- since we usually assume our theories are either  approximations to underlying, `effective' behaviour, or have certain simplifying assumptions ( for example expansion around equilibrium), we can assume the values of the constants in our theories are not always the exact physical ones, even if they are very close to them. Since our physical theories give good answers within their range of applicability, they should not be sensitive to small changes in these constants, including the structure constants. Hence the Lie algebras in physical theories should be stable. Fascinatingly, the transition from classical mechanics to both special relativity and quantum mechanics can be understood by going from an unstable to a stable algebra.\\
This is easiest to see in the case of special relativity.  Consider the Gallilean symmetry of Classical Mechanics. The algebra for the homogenous Galileo group is given by
\begin{align}\label{def:Galalg}
	[J_{i}, J_{j}]&=i\epsilon_{ijk}J_{k}\\
	[J_{i}, K_{j}]&=i\epsilon_{ijk}J_{k}\\
	[K_{i}, K_{j}&]=0
\end{align}
where the $J_{i}$ generate spatial rotations, and the $K_{i}$ are Galilean boosts. This algebra is unstable, but we can deform it to the Lorentz Algebra
\begin{align}\label{def:Lorentzalg}
	[J_{i}, J_{j}]&=i\epsilon_{ijk}J_{k}\\
	[J_{i}, K_{j}]&=i\epsilon_{ijk}J_{k}\\
	[K_{i}, K_{j}&]=-i\frac{1}{c^{2}}\epsilon_{ijk}K_{k}
\end{align}
This is the rotation part of the Poincare algebra,the symmetry group of special relativity and this is a stable algebra. Note that the deformation is implemented by a new parameter, the speed of light, $c$. \\
The transition from classical to quantum mechanics has additional technicalities (see \cite{Mendes1994}), but involves going to from the Poisson to the Moyal algebra. This induces a deformation from the classical position-momentum algebra $[X_{a}, P_{b}]=0$ to the Heisenberg Algebra
\begin{align}\label{def:Heisalg}
	[X_{a}, P_{b}]=\hbar\delta_{ab}
\end{align}
 with all other commutators being zero. Note, again, that making the deformation involves the introduction of a parameter, $\hbar$. \\
We might expect a relativistic quantum system to satisfy both the Heisenberg and Poincare-Heisenberg algebras. The simplest way to approach this is to take the direct product of the two algebras. This is called the Poincare- Heisenberg algebra \cite{Mendes1994}.  It is also the same algebra known in the literature \cite{Glikman2004} as, `doubly special relativity'. However, it turns out that this algebra is also unstable. The simplest stabilisation of this algebra was discovered by Mendes \cite{Mendes1994}\cite{Mendes1996} and is imaginatively known as the Stabilised Poincare-Heisenberg Algebra, or SPHA for short. It is given by 
\begin{align}\label{def:SPHA}
	[M_{ab}, M_{cd}]&=i\Big(M_{ad}\eta_{bc}+M_{bc}\eta_{ad}-M_{bd}\eta_{ac}-M_{ac}\eta_{bd}\Big)\\ \nonumber 
	[M_{ab}, P_{c}]& =i\Big(P_{a}\eta_{bc}-P_{b}\eta_{ac}\Big)\\ \nonumber 
	[M_{ab}, X_{c}]& =i\Big(X_{a}\eta_{bc}-X_{b}\eta_{ac}\Big)\\ \nonumber 
	[X_{a}, P_{b}]&=h\eta_{ab}I \\ \nonumber
	[X_{a}, X_{b}]&=-i\ell^{2}M_{ab}\\ \nonumber
	[P_{a}, P_{b}]&=-\frac{i}{R^{2}}M_{ab}\\ \nonumber
	[P_{a}, I]&=-\frac{i}{R^{2}}X_{a}\\ \nonumber
	[X_{a}, I]&=i\ell^{2}P_{a} \\ \nonumber
	[M_{ab}, I]& =0
\end{align}
where the $M_{ab}$ are the generators of the Lorentz group, $\eta_{ab}$ is the metric, $P_{a}$ are 4-momenta, $X_{a}$ are positions, and $l$ and $R$ are new deformation parameters whose interpretation is a subject of debate. II can choose the metric to have either Lorentzian or Euclidean signature. Along with $c$, which has here been set to $1$, there are three parameters (treating $h$ as a function of $l$ and $R$, usually taken as $h=l/R$) .  However, it is not a new algebra. It has also made appearances in other guises-- first, as a putative algebra for noncommutative spacetime in Yang \cite{Yang1947} and Snyder's \cite{Snyder1947} papers on quantized spacetime in the late 1940s, and also as the, `triply special relativity' put forward by Smolin in \cite{Smolin2004}. A discussion of these other versions can be found in \cite{Okon2004} \\
For our purposes, the most relevant property is that it is isomorphic to either the Spacetime Algebra  $Cl(3,1)$ or to $Cl(4)$, depending on whether we choose a Lorentzian or Euclidean metric. This was shown by Gresnigt et. al in \cite{Gresnigt2007}\footnote{Technically, that paper shows the isomorphism for $Cl(1,3)$ not $Cl(3,1)$}. I provide here a simpler mapping. If we compare the Spacetime Algebra, with commutators written as 
\begin{align}\label{def:SPHAGA}
	[\gamma_{ab}, \gamma_{cd}]&=\gamma_{ad}\eta_{bc}+\gamma_{bc}\eta_{ad}-\gamma_{bd}\eta_{ac}-\gamma_{ac}\eta_{bd} \\ \nonumber 
	[\gamma_{ab}, P_{c}]& =P_{a}\eta_{bc}-P_{b}\eta_{ac} \\ \nonumber 
	[\gamma_{ab}, X_{c}]& = X_{a}\eta_{bc}-X_{b}\eta_{ac} \\ \nonumber 
	[X_{a}, P_{b}]&=\hbar\eta_{ab}I \\ \nonumber
	[X_{a}, X_{b}]&=-\ell^{2}\gamma_{ab}\\ \nonumber
	[P_{a}, P_{b}]&=-\frac{1}{R^{2}}\gamma_{ab}\\ \nonumber
	[P_{a}, I]&=-\frac{1}{R^{2}}X_{a}\\ \nonumber
	[X_{a}, I]&=\ell^{2}P_{a} \\ \nonumber
	[\gamma_{ab}, I]& =0
\end{align}
with the SPHA in (\ref{def:SPHA}), we can see that they are identical, if we identify the central element $I$ in the SPHA with the pseudoscalar, or volume element, $I$ in the Spacetime Algebra. Here we are setting $\gamma_{a}=X_{a}$ and $\star\gamma_{a}=P_{a}$, a grade 3 object. \\
All this suggests that the Spacetime Algebra should be an important object for theoretical physics, since it incorporates the symmetries of quantum mechanics and special relativity. It is both the natural algebra induced on Minkowski space, and the simplest stable algebra related to the direct product of the Heisenberg and Minkowski algebras. Note too that positions and momentas are mapped to the geometrically complimentary grade 1 and grade 3 subspaces. This suggests there is a phase space structure hidden in the SPHA/ spacetime algebra and in $Cl(4)$. I will now use this suggestion to look for phase space stucture on Euclidean Clifford algebras in general. 
\section{Kähler Structure on Clifford Algebras}
	Recall that a Kähler Structure is defined by  
	\begin{equation}
		g(u,v)=\omega(u,Jv)
	\end{equation}
where $\omega$ is an antisymmetric two-form, $g$ is a metric, and $J$ is a complex structure satisfying $J^{2}=-1$.  In this section I present the main argument of this paper, that a Clifford Algebra can be equipped with a (quasi-) Kähler structure via the Hodge relation
\begin{equation}\label{eqn:CliffordKähler}
	g(A,B)\mathcal{I}=A\wedge\star B
\end{equation}
where $A$ and $B$ are grade $k$ elements within the Clifford Algebra. Why does it make sense to consider this as a generalisation of Kähler structure? First, the Hodge relation links the inner and outer products on the Clifford algebra, just as the usual Kähler structure links the symmetric and antisymmetric forms $g$ and $\omega$. Second, the Kähler stucture relates the element $v$ picked out by the metric, and $Jv$ picked out by the symplectic form. These have the same magnitude, but are orthogonal. The Hodge structure associates the element $B^{\dagger}$ picked out by the inner product with the element $\star B=B^{\dagger}I$. By definition,$B$ and $\star B$ are orthogonal-- they are an orthogonal decomposition of the vector space $\mathcal{V}$. We can also calculate
\begin{equation}
g(\star B, \star B)=\big(\star B\big)^{\dagger}\cdot\star B=\big(B^{\dagger}I\big)^{\dagger}\cdot B^{\dagger}I=I^{\dagger}B\cdot B^{\dagger}I=I^{\dagger}I\lvert B\rvert^{2}=\lvert B\rvert^{2}
\end{equation} 
Therefore in the euclidean case, these elements too have the same magnitude but are orthogonal. In the standard Kähler case, the elements are orthogonal vectors, here they are orthogonal subspaces. I therefore suggest that this is a generalisation of the Kähler structure from the Complex case to the Clifford case. \\
Why is this a generalisation of Kähler structure and not simply another example? The most clear reason is that $J^{2}=-1$ but $\star\star B=-1^{k(n-k)}B$. Therefore this is not an almost complex structure, but the sign depends both on the overall dimension $n$ and the dimension $k$ of the subspace represented by $B$. The second reason is that, whilst $g(A,B)$ is still symmetric, we have 
\begin{equation}
	A\wedge \star B = (-1)^{k(n-k)}\star B\wedge A
\end{equation}
so the outer product is not antisymmetric in general.
In my view these differences are due to the move from a pair of vectors to a pair of subspaces, and are an extension of Kähler structure rather than contradicting the possibility of such an extension-- but the overall justification for for seeing this as a generalisation of Kähler  structure will be its use in extending results from 2d Kähler geometry to the higher dimensional case. With this in mind, I will next examine how we can use this structure to define Hamiltonian dynamics on Euclidean Clifford algebras. \\
\subsection{Clifford Algebras as Phase Spaces}
Since the Clifford Algebra can be equipped with a Kähler structure, we can view it as a symplectic manifold, and hence as a phase space. The first thing to note is that we have a decomposition of a Clifford Algebra $Cl(n)$, into pairs of dual subspaces, which differs according to whether $n$ is odd or even. If $n$ is odd, we have
\begin{equation}\label{eqn:oddsplit}
	Cl(n)=\sum_{k=0}^{(n-1)/2} \mathcal{C}_{k}\otimes\mathcal{C}_{n-k}
\end{equation}
whereas if $n$ is even, we have
\begin{equation}\label{eqn:evensplit}
	Cl(n)=\sum_{k=0}^{n/2-1} \mathcal{C}_{k}\otimes\mathcal{C}_{n-k}+\mathcal{C}_{n/2}
\end{equation}
In the latter case, the, `middle' grade $\mathcal{C}_{n/2}$ splits into two non-unique dual parts. For example, in the case of $Cl(4)$, these are the grade 2 elements, spanned by $\{\gamma_{01}, \gamma_{02}, \gamma_{03}, \gamma_{12}, \gamma_{31}, \gamma_{23}\}$. This has many splittings into two dual spaces;for example we could take  the rotations involving time $\{\gamma_{01}, \gamma_{02}, \gamma_{03}\}$ and the purely spatial rotations $\{\gamma_{12}, \gamma_{31}, \gamma_{23}\}$, linked by $\gamma_{0i}=I\epsilon_{ijk}\gamma_{i}\gamma_{j}$. \\
Recalling that in the vector case, we have (locally) $\omega=\sum_{i} x_{i}\wedge p_{i}$. 
Recall also that for the SPHA in equation (\ref{def:SPHAGA}) we have the relation
\begin{equation}
X_{a}\wedge P_{b} = \hbar\eta_{ab}\mathcal{I}
\end{equation}
This implies the total volume should have units of energy. Therefore, we shall make the ansatz that the volume is given by the Hamiltonian function 
\begin{equation}
H(x,p)\mathcal{I} 
\end{equation}
 and the relation between the positions and the momenta is given by
  \begin{equation}
 P_{a}=X_{a}^{-1} I\equiv \star X_{a}/\lvert X_{a}\rvert^{2}
  \end{equation}
Using the Hodge relation, this guarantees that
\begin{equation}
	X_{a}\wedge P_{a}=X_{a}\wedge X_{a}^{-1}I=g(X_{a}, X_{a}/\lvert X_{a}\rvert^{2})I=I
\end{equation}
as required. From now on we shall work in units where $ \hbar=1$. \\ 
If we are regarding the grade $k$ subspaces $\mathcal{X}_{K}$ as associated with generalised positions $X_{k}$, and the $n-k$ grade subspaces $\mathcal{P}_{k}$  as associated with the corresponding momenta, denoted $P_{k}$,  then the splittings in equations (\ref{eqn:oddsplit}) and (\ref{eqn:evensplit}) become
\begin{equation}
	Cl(p,q)=\sum_{k=0}^{(n-1)/2} \mathcal{X}_{k}\otimes\mathcal{P}_{k}
\end{equation}
in the odd case, and 
\begin{equation}
	Cl(p,q)=\sum_{k=0}^{n/2-1} \mathcal{X}_{k}\otimes\mathcal{P}_{k}+\tilde{\mathcal{X}}_{n/2}\otimes\tilde{\mathcal{P}_{n/2}}
\end{equation}
in the even case, where $\tilde{\mathcal{X}}_{n/2}\otimes\tilde{\mathcal{P}_{n/2}}$ comes from the splitting of $\mathcal{C}_{n/2}$ in equation (\ref{eqn:evensplit}).  \\
We find that the pairs of variables $X_{k}, P_{k}$either commute or anticommute. Using  $P_{k}=X_{k}^{-1}I$, we find, for $Cl(p,q)$ with $p+q=n$:
\begin{equation}\label{eqn:anticomdef}
	 \big[X_{k}, P_{k}\big]=
	\begin{cases}
		0 & \frac{1}{2} k(n-k) \ even\\
		I & \frac{1}{2}k(n-k) \ odd
	\end{cases}
\end{equation}
 \begin{equation}\label{eqn:comdef}
	\big\{X_{k}, P_{k}\big\}=
	\begin{cases}
		I & \frac{1}{2} k(n-k) \ even\\
		0 & \frac{1}{2}k(n-k) \ odd
	\end{cases}
\end{equation}
Therefore for a fixed $n$, the $X_{k}, P_{k}$ pairs alternate between commuting (bosonic) and anticommuting (fermionic/ Grassman) behaviour. \\ 
To look at some common examples: For the Pauli Algebra $Cl(3)$=Span$\Big\{1, e_{i}, e_{ij}\Big\}$, we have $\mathcal{X}_{1}=\text{Span}\{e_{i}\}$ and $\mathcal{P}_{1}=\text{Span}\{e_{j}e_{k}\}$, which anticommute. For $Cl(4)$ we have:
	\begin{align}
		Cl(4)&=\text{Span}\Big\{\mathbb{1}, \gamma_{a}, \gamma_{ab}, \gamma_{abc}, I\Big\}\\
		&=\mathbb{1}+\mathcal{X}_{1}\otimes \mathcal{P}_{1}+\mathcal{X}_{2}\otimes \mathcal{P}_{2}
	\end{align}
 with
	\begin{align}\nonumber 
		\mathcal{X}_{1}&=\text{Span}\{\gamma_{a}\}\\ \nonumber 
	 \mathcal{P}_{1}&=\text{Span}\{\gamma_{a}\gamma_{b}\gamma_{c}\}\\ \nonumber 
 	\mathcal{X}_{2}&=\text{Span}\{\gamma_{i}\gamma_{j}\}\\ \nonumber 
		 \mathcal{P}_{2}&=\text{Span}\{\gamma_{0}\gamma_{i}\}
	\end{align}
 variables  $(X_{1}, P_{1})\in (\mathcal{X}_{1}, \mathcal{P}_{1})$ commute, and $X_{2}, P_{2}\in(\mathcal{X}_{2}, \mathcal{P}_{2})$ anticommute. 
	\subsection{Hamiltonian Mechanics on Clifford Algebras}
	Now we have our phase space structure, we can use it to define Hamiltonian mechanics. For simplicity I will consider a single $(\mathcal{X}_{k}, \mathcal{P}_{k})$ pair of subspaces (of grades $k$ and $n-k$), and with variables $X_{k}=x_{k}e_{K}$ and $P_{k}=p_{k}\star e_{K}$.   The general case can be found by summing over all such pairs. We begin by defining a position Dirac operator 
	\begin{equation}
	\vd_{X_{k}}\equiv e^{\dagger}_{K}\partial_{x_{k}}
\end{equation}
 and a momentum Dirac operator 
 \begin{equation}
 \vd_{P_{k}}\equiv (\star e_{K})^{\dagger}\partial_{p_{k}}=I^{\dagger} e_{K}\partial_{ p_{k}}
 \end{equation}
 Now, for a grade $k$ position $\vect{Q}_{k}=e_{K}Q_{k}$, and a grade $n-k$ momentum $\vect{P}_{k}=e_{K}^{\dagger}IP_{k}$, where $e_{K}$ is the grade $k$ basis, we get 		
	\begin{align}\nonumber
			\vd_{P_{k}}\cdot HI=&  I^{\dagger}e_{K}\frac{\partial H(x_{k},p_{k})}{\partial p_{k}}I =(-1)^{k(n-k)}e_{K}\frac{dx_{k}}{dt}=(-1)^{k(n-k)}\dot{X_{k}}\\ 
			\vd_{X_{k}}\cdot HI=&e_{K}^{\dagger}\frac{\partial}{\partial q_{k}}HI=-\frac{\partial H(x_{k},p_{k})}{\partial x_{k}}e_{K}^{\dagger}I
			=\dot{P_{k}}
		\end{align}
 These depend on the sign of $(-1)^{k(n-k)}$,  and hence on whether the pair commutes or anticommutes. If they commute, equations (\ref{eqn:comdef}) and (\ref{eqn:anticomdef}) show the sign is -, whereas if they anticommute then the sign is +. Comparing to (for example) the discussion in Henneaux and Teitelboim \cite{Henneaux1994}, we find that we have the correct  Hamiltonian equations for c-number (commuting) and Grassman (anti-commuting) variables. \\
	Finally, we can calculate Poisson brackets of scalar functions $F, G$. For simplicity we again focus the same $(\mathcal{X}_{k}, \mathcal{P}_{k})$ pair of subspaces (of grades $k$ and $n-k$) as above. We start by noting that by equation (\ref{eqn:poissonomega}) we might expect $\big\{F, G\big\}$ to be related to the wedge product of the derivatives of $F$ and $G$. I will therefore adopt the ansatz that 
	\begin{align} \nonumber
		\big\{F, G\big\}^{\dagger}& \equiv  \Big(\vd_{X_{k}}F+\vd_{P_{k}}F\Big)\wedge\Big(\vd_{X_{k}}G+\vd_{P_{k}}G\Big)\\ \nonumber 
		&=\vd_{X_{k}}F\wedge\vd_{P_{k}}G+\vd_{P_{k}}F\wedge\vd_{X_{k}}G\\ \nonumber 
		&=(F_{x_{k}}G_{p_{k}})e_{K}^{\dagger}\wedge e_{K}I^{\dagger}+(F_{p_{k}}G_{x_{k}})e_{K}I^{\dagger}\wedge e_{K}^{\dagger}\\
		&=\big(F_{x_{k}}G_{p_{k}}+(-1)^{k(n-k)}F_{p_{k}}G_{x_{k}}\big)I^{\dagger}
	\end{align}
Therefore
\begin{equation}
\big\{F, G\big\}=\big(F_{x_{k}}G_{p_{k}}+(-1)^{k(n-k)}F_{p_{k}}G_{x_{k}}\big)I
\end{equation}
Again, the sign $(-1)^{k(n-k)}$ is $-$ for $n$ even and $k$ odd, and $+$ otherwise, and once more this means the sign in the Poisson Bracket is $+$ for anticommuting (Grassmann) Variables and $-$ for commuting variables otherwise, exactly as in Henneaux and Teitelboim \cite{Henneaux1994}. This confirms that the $X_{k}, P_{k}$ pairs really are alternating bosonic and fermionic variables. This merits further investigation.  \\
Finally we can look at the Poisson bracket with the Hamiltonian. Replacing G with H in the above expression we have
\begin{align}\nonumber 
	\big\{F,H\big\}&=\big(F_{x_{k}}H_{p_{k}}+(-1)^{k(n-k)}F_{p_{k}}H_{x_{k}}\big)I\\ 
	&=\big(F_{q}\dot{q}-(-1)^{k(n-k)}F_{p}\dot{p}\big) I
\end{align}
Again, when we have commuting Euclidean variables this gives the expected result
\begin{equation}
	\big\{F, H\big\}=F_{q}\dot{q}+F_{p}\dot{p}
\end{equation} 
When we have anticommuting Euclidean variables we get 
$\{F, H\}=F_{q}\dot{q}-F_{p}\dot{p}$ with the same relative minus as above. 
	\section{Conclusion}
	This shows that we can define Hamiltonian dynamics from the Hodge relation, just as we can from the standard Kähler structure. This suggests that the analogy between them is genuine, and that (just as with other parts of complex analysis) we can generalise Kähler structure to higher dimensions via Clifford algebras. The Kähler structure is generalised from orthogonal vectors to orthogonal subspaces. The Hamiltonian dynamics from the Hodge relation are generalised from the Kähler case in that we can describe $k-$ and $(n-k)-$ dimensional pairs of subspaces rather than pairs of vectors. \\
There are four main directions to explore here. First, there is the curious fact that the paired position/momentum subspaces are alternating commuting and anticommuting variables. This feature of the analysis was unexpected, and seems further evidence that the construction in this paper is a natural one. Studying this behaviour further and trying to find examples of concrete physical theories incorporating them would be very worthwhile. Second, there is the fact that we have here implicitly considered the generalised Kähler structure on a flat background.  Implicitly, the we have an, `almost Hodge structure' on any tangent space equipped with a Clifford algebra, but in the  The next step, therefore, is to extend the analysis here to curved manifolds.  Thirdly it would be instructive compare the noncommutative phase space structure in this paper with the quantum phase space associated with Moyal\cite{Moyal1949}\cite{Hiley2012}. Finally, I intend to extend the ideas here to the non-euclidean case. 
	\bibliographystyle{siam}
	\bibliography{mathspapers2025}
\end{document}